\pdfoutput=1
\documentclass[preprint,12pt]{elsarticle}

\usepackage{amssymb}
\usepackage{amsmath}
\usepackage{url}
\usepackage{lineno}
\usepackage[colorlinks=true, linkcolor=blue, citecolor=blue, urlcolor=blue]{hyperref}

\journal{Nuclear Physics B}

\begin{document}

\begin{frontmatter}


\title{Concept of the UCN Source at the WWR-K Reactor (AlSUN)}

\author[label1]{Sayabek Sakhiyev} 

\author[label1,label3]{Kylyshbek Turlybekuly \corref{cor1}}
\ead{k.turlybekuly@inp.kz}

\author[label1]{Asset Shaimerdenov}
\author[label1]{Darkhan Sairanbayev}
\author[label1]{Avganbek Sabidolda}
\author[label1,label3]{Zhanibek Kurmanaliyev}
\author[label1]{Akzhol Almukhametov}
\author[label1]{Olzhas Bayakhmetov}
\author[label1]{Ruslan Kiryanov}
\author[label2]{Ekaterina Korobkina}
\author[label3]{Egor Lychagin}
\author[label3]{Alexey Muzychka}
\author[label4]{Valery Nesvizhevsky}
\author[label5]{Cole Teander}
\author[label6]{Pham Khac Tuyen }

\affiliation[label1]{organization={Institute of Nuclear Physics}, 
                     addressline={1 Ibragimov Street}, 
                     city={Almaty}, 
                     postcode={050032}, 
                     country={Kazakhstan}}

\affiliation[label2]{organization={Department of Nuclear Engineering}, 
                     addressline={North Carolina State University }, 
                     city={Raleigh}, 
                     postcode={NC 27695}, 
                     country={USA}}

\affiliation[label3]{organization={Frank Laboratory of Neutron, Physics Joint Institute for Nuclear Research}, 
                     addressline={6 Joliot Curie}, 
                     city={Dubna}, 
                     postcode={Ru-141980}, 
                     country={Russia}}    

\affiliation[label4]{organization={Institut Max von Laue-Paul Langevin}, 
                     addressline={1 Av. des Martyrs}, 
                     city={Grenoble}, 
                     postcode={F-38042}, 
                     country={France}}  
\affiliation[label5]{organization={Department of Physics}, 
                     addressline={North Carolina State University }, 
                     city={Raleigh}, 
                     postcode={NC 27695}, 
                     country={USA}}                       
\affiliation[label6]{organization={Moscow Institute of Physics and Technology (MIPT)}, 
                     addressline={North Carolina State University }, 
                     city={Dolgoprudny}, 
                     postcode={141700}, 
                     country={Russia}}     
                     
\begin{abstract}
We present the concept of the ultracold neutron (UCN) source with a superfluid helium-4 converter located in the thermal column of the WWR-K research reactor at the Institute of Nuclear Physics (INP) in Almaty, Kazakhstan. The conceptual design is based on the idea of accumulating UCNs in the source and effectively transporting them to experimental setups. We propose to improve the UCN density in the source by separating the heat and the UCN flows from the production volume and decreasing both, the temperature of the SuperFluid $^4$He (SF $^4$He) converter below $\sim$1 K and the coefficient of UCN wall loss below $\sim10^{-4}$. To achieve the operation temperatures below 1 K we plan to use a He-3 pumping cryogenic system and minimize the thermal load on the UCN accumulation trap walls. Additional gain in the total number of accumulated UCNs can be achieved due to use of a material of a high critical velocity for the walls of the accumulation trap. The implementation of such a design critically depends on the availability of materials with specific UCN and cryogenic properties. This paper describes the conceptual design of the source, discusses its implementation methods and material requirements, and plans for material testing studies.
\end{abstract}




\end{frontmatter}



\section{Introduction}
\label{sec1}

Ultracold neutrons (UCNs) were first experimentally discovered in 1968 at the Joint Institute for Nuclear Research (JINR) in Dubna by a group led by F. L. Shapiro \cite{Lushchikov1969}. Independently of these works, Albert Steyerl observed in 1969 UCNs in the low-energy tail of the thermal spectrum of a neutron source when measuring the total cross sections of interaction of low-energy neutrons with substances \cite{Steyerl1969}. Already in 1974, the first experiments with UCNs were carried out at the WWR-K reactor at INP in Almaty, Kazakhstan \cite{Akhmetov1974, Akhmetov1977}.

UCNs play an important role in fundamental research. Due to their extremely low kinetic energies, and hence long wavelengths, they can be stored for a long time in traps, making them an ideal tool for making precise and/or sensitive measurements, such as measuring the lifetime and asymmetry coefficients of neutron $\beta$-decay, searching for the electric dipole moment and charge of the neutron, measuring gravitational and whispering-gallery quantum states of neutrons, in particular for searching for additional fundamental interactions and testing fundamental symmetries \cite{Ignatovich1990,Golub1991, Nesvizhevsky2002,Bertolami2005,Lamoreaux2009, Liu2010,Nesvizhevsky2010,Antoniadis2011, Dubbers2011, Serebrov2011, Pedram2011,Engel2013, Jenke2014, Siemensen2015, Greene2016, Abel2020,Gonzales2021,Dubbers2021, Fomin2022, Guerrero2024}.

The intensity of existing UCN sources is constantly improving. For instance, density of 273 UCN per cc in the accumulation volume of superSUN source has been reported recently in~\cite{superSUN2025}. Nevertheless, for high precision experiments data taking still takes years and many experiments are statistically limited. To overcome this limitation and expand the range of possible fundamental and applied research, further technical development needed to obtain even higher UCNs statistics. One of the most promising is the use of SuperFluid $^4$He (SF $^4$He) as a converter, the operating principle of which is based on the conversion of Cold Neutrons (CNs) into UCNs as a result of single inelastic processes of their scattering on collective excitations in SF $^4$He (phonons).
Several neutron centers around the world are developing and improving UCN sources based on SF $^4$He converters. There are two main options for the location of the converter: close to the neutron source (for example, next to the reactor core) and on the extracted beams of cold neutrons. Since the density of UCNs is proportional to the flux of incident neutrons, it is preferable to place the converter close to the neutron source, as well as to ensure the most complete (close to $4\pi$) solid angle of irradiation of the converter.

In Section 2 we propose the concept of the UCN source for the WWR-K reactor, Almaty (AlSUN: Almaty Source of Ultracold Neutrons) based on the development of existing projects (TRIUMF, PNPI). In Section 3, possible parameters of the AlSUN source are estimated. Section 4 discusses the problems that must be solved in order to cool the converter to the required temperatures, to select the material for the walls of the UCN accumulation trap, and to efficiently deliver UCNs from the source to the experimental setups.

\section{AlSUN source concept}
\label{sec2}
\subsection{Development of the method of UCN converters based on SF $^4$He}
\label{subsec2}

In 1975, Golub and Pendlebury proposed using collective excitations in SF$^4$He to produce UCNs \cite{Golub1975}. The key insight is to exploit the collective excitations of SF$^4$He, particularly single-phonon emission at a neutron wavelength corresponding to an energy of 8.9~\AA~. This process allows neutrons to lose almost all their kinetic energy in a single quantum event when their wavelength aligns with the SF $^4$He phonon spectrum. The mechanism is most effective for neutrons with wavelengths around 8.9~\AA~, while neutrons with wide spectrum can still produce UCNs through less efficient multi-phonon emission. At low temperatures, pure SF $^4$He exhibits extremely favorable conditions for UCN storage. Due to its zero neutron absorption cross section, SF $^4$He does not capture neutrons, making it an ideal medium to preserve neutron populations. Furthermore, at very low temperatures, the density of collective excitations in SF $^4$He becomes vanishingly small. This significantly suppresses the probability that the UCN is undergoing inelastic scattering back to higher energies. As a result, UCN can be stored for extended durations - potentially several hundreds of seconds - provided the converter vessel is constructed from materials with a very low probability of UCN loss due to absorption or wall interactions. This combination of minimal neutron absorption and low excitation density establishes SF $^4$He as one of the most efficient environments for the production and storage of UCN. Consequently, a very high density of UCNs was predicted in a converter cooled to temperatures below $\sim$1 K. In works \cite{Golub1983, Yoshiki1992} the possibility of producing UCNs in SF $^4$He was experimentally shown.

As an example of the location of an SF $^4$He converter, a large channel leading to the "Bulk Shielding Experimental Tank (BSET)" of the TRIGA Mark II reactor in Vienna, Austria was considered \cite{Golub1984}. For this reactor, the authors proposed the following concept of an UCN source: Graphite moderator at room temperature slows fission neutrons down to thermal energies; SF $^4$He at a temperature below $\sim$1~K converts thermal energy neutrons into UCNs; a 12~cm thick bismuth gamma screen reduces the thermal load on the SF $^4$He. 

According to the authors, the installation of such an UCN source in a large channel leading to the BSET Mark II TRIGA reactor can provide the source parameters listed in Table \ref{tab:TRIGA}.

\begin{table*}[t]
    \centering    
    \begin{tabular}{l c c r} 
    \hline
     Neutron         & TN flux,     & Thermal load,  & UCN density, \\
     source & $n/cm^2/s$ & mW & $UCN/cm^3$ \\
    \hline
     TRIGA Mark II (250 kW) & $8\cdot 10^{10}$ & 30  & 200 \\
    \hline
    \end{tabular}
    \caption{Parameters of the proposed UCN source at the TRIGA reactor.}\label{tab:TRIGA}    
\end{table*}

Thus, the work \cite{Golub1984} proposed for the first time the concept of an UCN source with a SF $^4$He converter, located near the reactor core and protected by a $\gamma$-quanta screen to reduce the thermal load. The main advantages of this concept are the use of the $4\pi$ angular distribution of incident neutrons and the efficient accumulation of UCNs due to the decrease in the temperature of SF $^4$He.

Inspired by this concept, a group of researchers from Japan proposed placing a SF $^4$He converter near the spallation source \cite{Masuda2000}. According to this idea, fast neutrons generated by the interaction of a 600~MeV proton beam with a 20~$\mu$A current with a target are slowed down to cold energies by a two-step moderation: first in heavy water (D$_2$O) at room temperature, and then in liquid deuterium (LD$_2$) at a temperature of $\sim$20~K. UCNs are produced in a cryogenic volume with SF $^4$He cooled to a temperature of $\sim$0.5~K and placed in a zone with a high neutron flux density. To protect against $\gamma$-quanta coming from the target, a bismuth (Bi) screen $\sim$10~cm thick is installed. This idea was subsequently implemented at the Nuclear Physics Research Center of Osaka University in Japan using a spallation source (400~MeV and 1~$\mu$A current) \cite{Masuda2002}.

After experimentally confirming the cooling of a volume with SF $^4$He located near the target by pumping out helium-3 ($^3$He) vapor, a group of scientists from Japan and Canada decided to place such an UCN source in a neutron source based on splitting the tungsten target with protons with an energy of 483~MeV and a power of 19.3~kW at the TRIUMF accelerator complex \cite{Masuda2012,Ahmed2019}. As a result of extensive modeling and optimization of the source geometry, the expected UCN production rate was estimated to be $P_{UCN}\sim (1.4-1.6)\cdot 10^7~UCN/s$, and the UCN density in the $70~l$ experimental setup $\rho_{UCN}\sim 220~UCN/cm^3$ \cite{Schreyer2020,Sidhu2023}. Up to $\sim$10~W of heat is generated in the UCN source. This heat is removed through a channel, 2.5~m long and 0.15~m in diameter, filled with SF $^4$He. The same channel is used to extract UCNs. Then the heat is transferred to the cooling system, which is based on pumping out $^3$He vapors.

A similar concept of an UCN source was later proposed for the WWR-M research reactor with a power of 16~MW by the group led by Serebrov (PNPI, Russia) \cite{Serebrov2015, Serebrov2016, Onegin2017}. To implement it, the authors proposed, as in the original design by Golub, using a large-diameter channel (100~cm), the thermal column of the WWR-M reactor, adjacent to the reactor core. 

According to the authors, such a parameter of the reactor thermal column allows the proposed concept to be implemented and the problem of maintaining a balance between the thermal load and the neutron flux to be solved. The reactor thermal column can accommodate lead (Pb) screen from $\gamma$-quanta of the reactor core, a graphite moderator and a LD$_2$ premoderator at a temperature of $\sim$20~K to produce CNs, as well as a chamber with SF $^4$He converter at a temperature of $\sim$1.2~K. The authors assumed that the UCN production would be $P_{UCN}\sim (6-8)\cdot 10^7~UCN/s$, and the UCN density in an experimental setup with a volume of $35~l$ would be $\rho_{UCN}\sim 10^4~UCN/cm^3$.

These two similar projects differ in the types of neutron sources: one uses a spallation reaction, and the other a fission reaction. In addition, their cooling systems for the SF $^4$He converter differ. In the first project, cooling is achieved by pumping out $^3$He vapor, which theoretically allows temperatures of $\sim$0.6~K to be reached. In the second project, $^4$He vapor is used, which allows temperatures to be reached of only down to $\sim$1.2~K.

The location of SF $^4$He near the active zone has also been proposed for the Los Alamos Neutron Center (Santa Fe) \cite{Leung2019}, as well as for small-size low-energy proton accelerators \cite{Shin2021}. Sources operating on extracted neutron beams have been implemented, for example, at the high-flux reactor of the Laue-Langevin Institute (Grenoble) \cite{Piegsa2014, Chanel2022}. A compromise solution is considered in the work \cite{Lychagin2016}. It provides a large total neutron flux and 4$\pi$ angular distribution at an extracted TNs/CNs beam. 

\subsection{Concept of the UCN source at the WWR-K reactor}
\label{subsec21}

\begin{figure*}[t]
\centering

\includegraphics[width=0.7\textwidth]{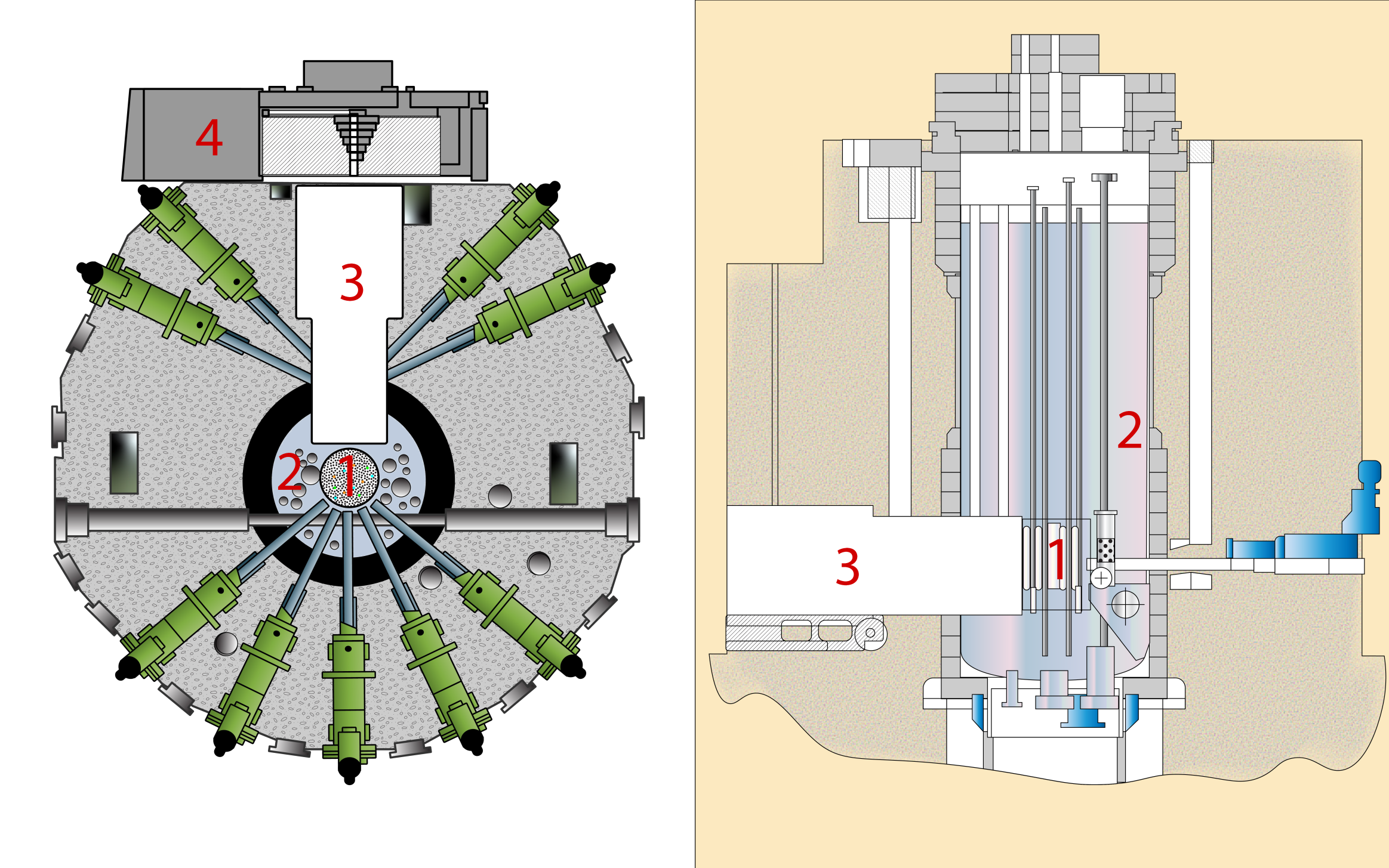}
\caption{Scheme of the WWR-K reactor, top view (left) and side view (right). 1 – reactor core; 2 – moderator, H$_2$O; 3 – thermal column; 4 – rollback protection.}\label{fig:WWR-K}
\end{figure*}

The research reactor WWR-K (Water-Water Reactor - Kazakhstan) at INP \cite{Shaimerdenov2018} has a power of 6~MW, while its design is almost no different from the WWR-M reactor (Gatchina), which had a power of 16~MW. The difference in reactor power is explained by the fact that the WWR-K reactor uses safer low-enriched fuel. As in the case of the WWR-M reactor, the WWR-K reactor has a thermal column, which is a large-diameter channel (1 meter) adjacent to the reactor core (Fig. \ref{fig:WWR-K}).

\begin{figure*}[t]
\centering
\includegraphics[width=0.8\textwidth]{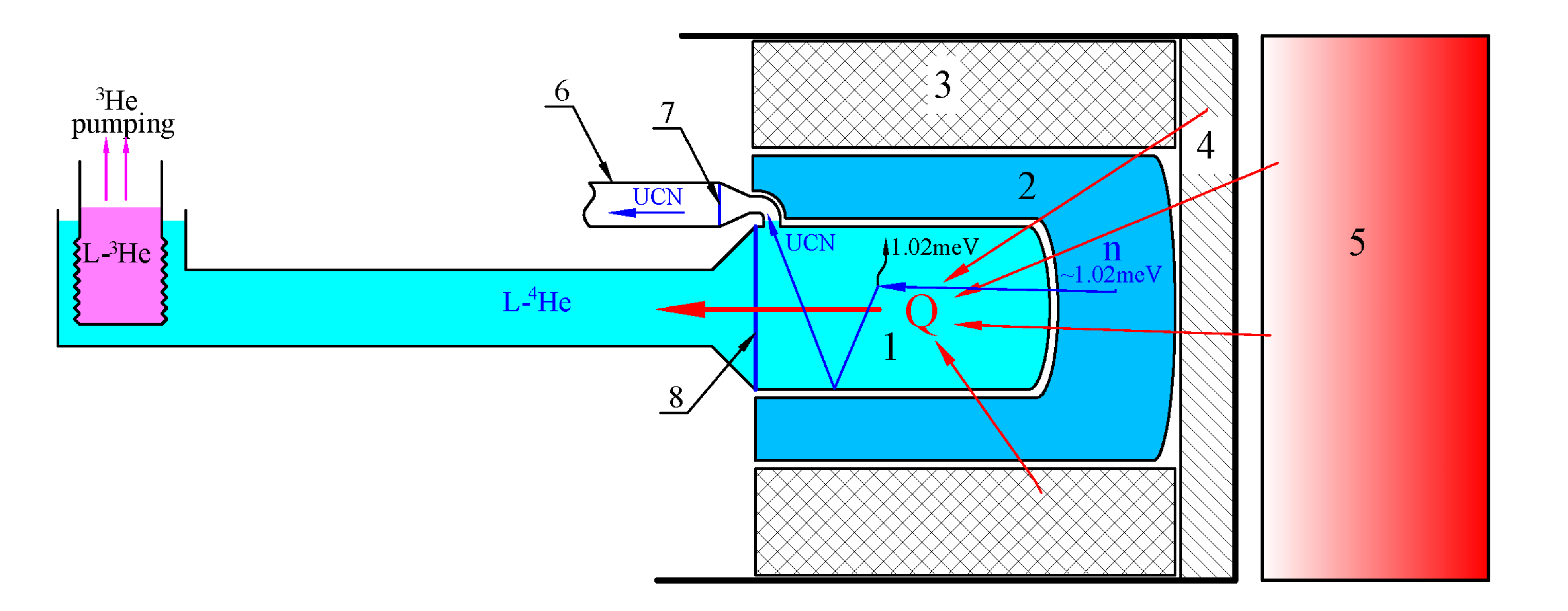}
\caption{Scheme of the UCN source at the WWR-K reactor.
1 – trap with SF $^4$He at a temperature T<1~K; 2 – LD$_2$, T$\sim$20~K; 3 – graphite, T$\sim$300~K; 4 – Pb; 5 – reactor active zone; 6 – neutron guide; 7 – separating foil; 8 – heat-conducting wall.}
\label{fig:SourceScheme}
\end{figure*}

Neutrons enter the thermal column primarily through its front wall. The calculation performed using the MCNP6.2 program \cite{Goorley2013} shows that the maximum flux density of TNs on the front wall of the thermal column is $1.21\cdot 10^{12}~cm^{-2}s^{-1}$ \cite{Turlybekuly2024}. We plan to place the UCN source near the front wall of the thermal column, the diagram of which is shown in Fig. \ref{fig:SourceScheme}.

The main element of the source is a cylindrical trap with SF $^4$He (1, Fig. \ref{fig:SourceScheme}) at the temperature T<1~K, in which UCNs are formed from CNs and accumulated. CNs, in turn, are formed in LD$_2$ (2, Fig. \ref{fig:SourceScheme}) from TNs and fast neutrons. LD$_2$, which is one of the best cold moderators \cite{Ageron1989}, surrounds the trap (1, Fig. \ref{fig:SourceScheme}) with SF $^4$He on three sides and has a temperature T$\sim$20~K.

Fast neutrons are produced by nuclear fission in the reactor core (5, Fig. \ref{fig:SourceScheme}). The core is surrounded by a light water moderator (H$_2$O, not shown in Fig. \ref{fig:SourceScheme} for simplicity), which slows down fast neutrons to thermal energies. The water layer between the core and the front wall of the thermal column has a variable thickness with a minimum at 5~cm, in the central part of this wall . The maximum of the thermal neutron flux also enters the UCN source from the central part of the front wall of the thermal column. The graphite layer surrounding the LD$_2$ tank (3, Fig. \ref{fig:SourceScheme}) serves as an additional thermal neutrons moderator. 

A large flow of $\gamma$-quanta comes from the reactor core and the H$_2$O surrounding it, resulting in significant radiation heating of SF $^4$He. To reduce this heat influx, a Pb screen (4, Fig. \ref{fig:SourceScheme}) is installed close to the front wall of the thermal column, which reduces the flow of $\gamma$-quanta hundreds times. Although to a lesser extent, the lead shield also reduces the neutron flux, thereby reducing the productivity of the UCN source. Therefore, the thickness of the Pb screen must be optimized in such a way as to keep the total radiation heat within capacity of the cryogenic equipment while avoiding an excessive reduction in the neutron flux. Note, that a source capable of working in UCN accumulation mode might benefit from working at lower neutron fluxes with additional gamma-shielding if it allows cryogenic to get down to below 0.9 K operation temperatures.

The UCNs produced in the source can be reflected multiple times from the internal walls of the accumulation trap and remain there until they enter the exit located at the top of the trap (1, Fig. \ref{fig:SourceScheme}) or they are lost due to heating by phonons in SF $^4$He or due to losses in the source trap walls. Then they pass through a sealed separating foil (7, Fig. \ref{fig:SourceScheme}) and enter the neutron guide (6, Fig. \ref{fig:SourceScheme}), which leads to the experimental setups. The source walls must be coated with a material with a high critical energy and a low UCN loss coefficient to ensure multiple reflection of the UCNs from them.

The separating foil must be made of a material with a low critical energy and a small UCN loss cross-section (usually aluminum (Al) is used for this). The separating foil performs two functions: first, it reflects the black body radiation coming from the higher temperature neutron guide; second, it separates the vacuum of the neutron guide and the source, preventing cryopumping of contaminants by the SF $^4$He bulk.

The thickness of the LD$_2$ layer on the reactor side is $\sim$2 times greater than its thickness around the cylindrical surface of the trap (1, Fig. \ref{fig:SourceScheme}). This is necessary to prevent fast neutrons from entering the SF $^4$He, which are another factor in the influx of heat into the source.

The design of our source (Fig. \ref{fig:SourceScheme}) is similar to the design of the UCN source in the PNPI project, which is not surprising given the similarity of the designs of the reactors themselves. However, our source design has significant differences from both projects: PNPI and TRIUMF. In both of these projects, heat is removed due to free circulation of SF $^4$He between the UCN source and the cooling system, which is located 2-3~m from the source, behind the biological protection. SF $^4$He freely circulates in a wide tube, 14-15~cm in diameter, and there is a hole of the same diameter in the source wall. The same tube is also a neutron guide, through which UCNs are transported from the source to the experimental setups. Therefore, UCNs accumulate simultaneously in the entire source - neutron guide - experimental setup system. The UCN density is diluted proportionally to the increase in the total volume of the system. In addition, a long storage time of UCNs must be provided in each part of this system. It is most difficult to provide this in the neutron guide, which necessarily contains separating foils through which UCNs pass, repeatedly and in both directions. As a rule, neutron guides also contain movable devices with significant gaps (dampers, movable shutters, etc.).

The advantage of such a design is the reduced requirements for the UCN storage time in the source, and, consequently, for the SF $^4$He temperature in it, as well as for the quality of the accumulation trap walls. The productivity of the UCN source of such a design may even be somewhat higher than ours, due to a higher allowed limit of heat inflow, however, this factor may be leveled out due to large losses of UCNs during their transport.

In our source, we plan to separate the heat flow and the UCN flow. For this purpose, the heat is going to be removed through the heat-conducting wall (8, Fig. \ref{fig:SourceScheme}) of the source, and the UCNs are extracted through a small hole. Then the UCNs are transported to the experimental setups through a special focusing neutron guide, described in more detail in Section 4. The UCN density at its exit can be close to the UCN density at its entrance. This method provides the possibility of UCN accumulation to a high density both in the source and in the experimental setup.

To provide that the temperature of SF $^4$He in the source is less than $\sim$1~K, it is necessary to use a cooling system based on pumping out $^3$He vapor, which is similar to that used in the TRIUMF project and is schematically shown in Fig. \ref{fig:SourceScheme}.

The described concept of the UCN source contains new, untested methods, such as cooling the source through a heat-conducting wall to a temperature below $\sim$1~K and the use of focusing UCN neutron guides, so it is necessary to prove the feasibility of its implementation. For this purpose, preliminary studies are planned in three directions: 1. Measurement of the amount of heat that can be removed from SF $^4$He through a heat-conducting wall at a temperature of $\sim$1~K and below, depending on the temperature; 2. Study of surface coatings with high critical energy that can be used to cover the inner walls of the UCN source; 3. Calculation and development of focusing UCN guides.

\section{Results}
\label{sec3}
\subsection{Estimates of the 8.9~\AA~neutron fluxes}
\label{subsec31}

Using the MCNP6 program \cite{Goorley2013}, we calculated several possible UCN source geometries similar to the scheme shown in Fig. \ref{fig:SourceScheme}. These calculations had two goals. The first is to determine the dependence of the differential neutron flux density in the source on the neutron wavelength, $dJ/d\lambda(\lambda)$. Using it, we calculate the flux density at the wavelength $\lambda=8.9$~\AA, corresponding to the neutron energy of $\sim$1.02~meV. Using this value, we calculated the UCN source productivity neglecting multiphonon processes, bacause taking them into account only slightly increase the source productivity and the UCN density. The second goal is to calculate the heat influx to all elements of the source.

Several shielding and moderators geometry were simulated. Here, we present results for the AlSUN geometry which is very close to the PNPI project: the volume of the cylindrical trap with SF $^4$He (1, Fig. \ref{fig:SourceScheme}) is 35~l (diameter 30~cm, height~50 cm); the volume of LD$_2$ is $\sim$90~l, the LD$_2$ thickness on the reactor core side is 20~cm; the thickness of the Pb is 10~cm. Such choice allows us  to benchmark our calculations to eliminate potential errors and conservatively estimate expected UCN yield. Note that as was described above, the PNPI geometry was optimized for continuous UCN extraction without any accumulation of UCN in the source, i.e. for the short lifetime of UCN in the production volume due to a large diameter of UCN extraction guide. Nevertheless,  it is still a good first  approximation to the optimal geometry for our UCN source as well. Indeed, the optimization of accumulation mode requires a good quantitative knowledge of the actual radiative heating. Therefore,  it should be done with a more matured engineering design and realistic materials. 

\begin{figure}[t]
\centering
\includegraphics[width=1.0\columnwidth]{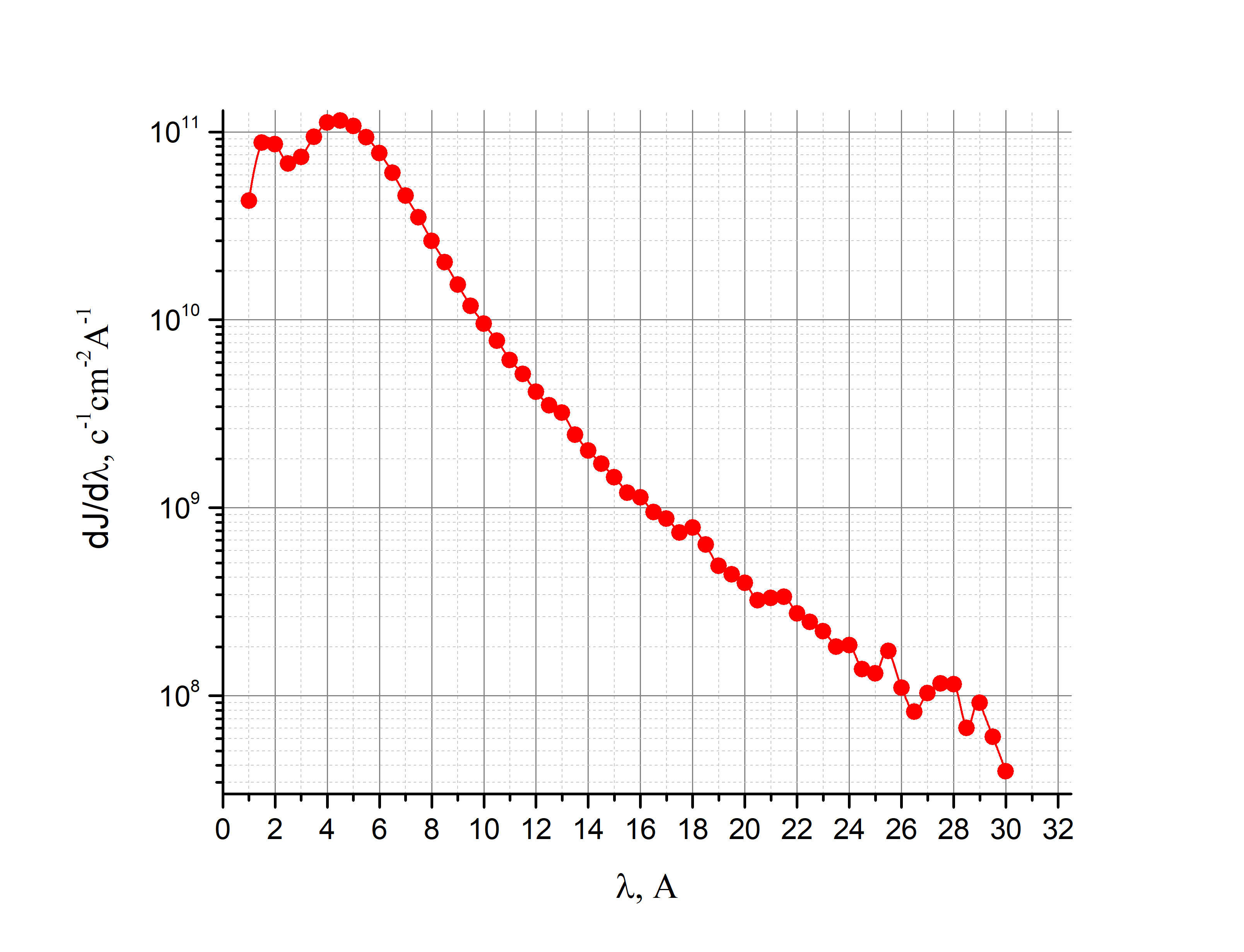}
\caption{Differential neutron flux density averaged over the volume of SF $^4$He trap.}
\label{fig:DifFlux}
\end{figure}

Figure \ref{fig:DifFlux} shows the dependence of the differential neutron flux density on the neutron wavelength in the trap with SF $^4$He, averaged over its volume.

The flux density of CNs with the wavelength of 8.9~\AA~($\sim$1.02~meV), extracted from this dependence, is:
$dJ/d\lambda(8.9~$\AA$)=1.62\cdot 10^{10}~cm^{-2}~s^{-1} $\AA$^{-1}$.

\subsubsection{Estimates of UCN production}
\label{subsec32}

The specific rate of production of UCNs with energies lower than the beryllium (Be) critical energy ($\sim$252~neV), $R$, due to the one-phonon process in 1~cm$^3$ of SF $^4$He can be calculated using formula $^{Be}R=4.55\cdot 10^{-8}\cdot dJ/d\lambda (8.9~$\AA$)~cm^{-3}s^{-1}$ \cite{Korobkina2002}. Here, the letter $R$ is preceded by the index $^{Be}$, which indicates that the part of the UCN spectrum limited by the Be critical energy is taken into account (UCNs are produced in a trap with Be-coated walls).

When UCNs are produced, they uniformly fill the phase space, so their spectrum will be as follows:

\begin{equation}
\label{eq:spectrum}
dR(E)/dE\sim \sqrt{E},
\end{equation}
where $E$ is UCN energy.

The rate of production of UCNs will depend on the critical energy (Fermi potential or optical potential) of the trap walls, $E_{lim}$, as follows:

\begin{equation}
    R(E_{lim})\sim (E_{lim}-E_{He})^{3/2},
\end{equation}
where $E_{He}\sim 18.5~neV$ is the critical energy of SF $^4He$.

The production rate of the UCN source with the volume of V (in our case $V=35~l$):

\begin{equation}
P_{UCN}=RV.
\end{equation}

For our source, with trap walls coated with Be, $^{Be}P_{UCN}=2.6\cdot 10^7~s^{-1}$, which is in good agreement with the calculations done at PNPI ($^{Be}P_{UCN}=(6-8)\cdot 10^7~ s^{-1}$) \cite{Onegin2017}, taking into account that the power of the WWR-K reactor is $\sim$3~times lower than the power of the WWR-M reactor.

Calculation of the heat load on the source showed that if the trap with SF $^4$He is made of Al with walls 2~mm thick, then the total heat load on the trap walls and He is $\sim$10~W at the 6~MW reactor power, with almost 2/3 of this heat coming from the Al. If the material of the trap wall, or a part of it, is replaced from Al to Be or zirconium (Zr), then the heat influx into the source can be significantly reduced. We plan to consider the possibility of using a Al-Be alloy or a zircaloy -IV.

Let us estimate the maximum UCN density that can be produced in the source. To do this, we consider a closed source (the UCN exit from the trap is closed).

\begin{equation}
\rho_{max}=\tau^0_{stor}\cdot R=\tau^0_{stor}\cdot P_{UCN}/V .
\end{equation}

The storage time of UCNs in the closed source, $\tau^0_{stor}$, is determined by the formula:

\begin{equation}
\frac{1}{\tau^0_{stor}}=\frac{1}{\tau_{He}}+\frac{1}{\tau_{wall}}+\frac{1}{\tau_\beta} ,
\end{equation}
where $\tau_{He}$ is the partial UCN storage time in SF~$^4$He, $\tau_{wall}$ is the partial UCN storage time in the trap determined only by losses during their interaction with the walls of the trap, and $\tau_\beta\sim 878~s$ is the partial neutron lifetime determined by its $\beta$-decay.

$\tau_{He}$, as indicated above, strongly depends on He temperature $\tau_{He}\sim T^{-7}$ \cite{Korobkina2002}. Let us list the values of $\tau_{He}$ at a few temperatures in Table \ref{tab:HeTimes}.

\begin{table}[]
    \centering
    \begin{tabular}{l c r}
    \hline
    He temperature, K & $\tau_{He},~s$ & $\tau^0_{stor},~s$ \\
    \hline
    1.2 & 35 & 27 \\
    1.1 & 70 & 44 \\
    1.0 & 130 & 62 \\
    0.9 & 260 & 82 \\
    0.8 & 610 & 100 \\
    \hline
    \end{tabular}
    \caption{Partial storage time of UCNs in SF $^4$He  and storage time of UCNs in the closed source with Be walls and $\eta =3\cdot 10^{-4}$ at different SF $^4$He temperatures.}
    \label{tab:HeTimes}
\end{table}

$\tau_{wall}$ is determined by the probability $\mu$ of UCN losses while they are reflected from the trap walls, and the frequency of impacts on the walls is $\nu_{wall}$: $\tau_{wall}=\mu/\nu_{wall}$. In turn, the probability of losses $\mu$ is determined by the loss factor $\eta$, which is the ratio of the imaginary and real parts of the Fermi potential of the wall material. The values of the loss factors $\eta$ for different materials vary greatly, while the theoretical values of $\eta$ are usually significantly lower than those achieved in practice. For example, the theoretical value for Be is $\eta\sim 3\cdot 10^{-7}$, and the minimum value obtained in practice is $\eta\sim 3\cdot 10^{-5}$ \cite{Alfimenkov1992}. In the calculations of TRIUMF and PNPI, the value $\eta =3\cdot 10^{-4}$ was chosen. We will also choose it.

Thus, we get an estimate of the storage time of UCNs in the closed source with Be walls and $\eta =3\cdot 10^{-4}$ at different He temperatures given in Table \ref{tab:HeTimes}.

If we manage to achieve a He temperature of $\sim$0.9~K, we get: $\rho_{max}(0.9~K)=6.1\cdot 10^4~UCN/cm^{3}$.

\subsection{Extraction and transport of UCNs}
\label{subsec33}

Let the UCN source have an exit hole of area $S_{exit}$, leading to a neutron guide from which neutrons do not return. Then the storage time in the source will be determined by the formula:

\begin{equation}
    \frac{1}{\tau_{stor}}=\frac{1}{\tau^0_{stor}}+\frac{1}{\tau_{exit}}.
\end{equation}
Emptying time is $\tau_{exit}=1/\nu_{exit}$, where exit frequency is $\nu_{exit}=\nu_{wall}\cdot S_{exit}/S_{wall}$, and $S_{wall}$ is the area of the source walls. In our case $S_{wall}=6126~cm^2$. Let $S_{exit}$ be such that $\tau_{exit}=\tau^0_{stor}(0.9~K)=82~s$, then $S_{exit}=3.6~cm^2$ at an UCN velocity of $5~m/s$. In this case, the total flux of UCNs exiting the source and entering the neutron guide is $F^0_{exit}=P_{UCN}/2$.

If an expanding mirror cone is installed at the beginning of a mirror neutron guide, the motion of UCNs will become more directed along the axis of the neutron guide. The frequency of impacts on the walls of the neutron guide, and, consequently, the losses of UCNs in it, will be sharply reduced. Moreover, the larger the ratio of the neutron guide diameter to the size of the source exit opening, the less often the UCNs will hit the walls. At the end of the neutron guide, before the entrance to the experimental setup, a tapering cone can be placed, focusing the UCNs onto an area equal to $S_{exit}$. This is the same principle that is used in the design of so-called ballistic neutron guides for CNs and TNs \cite{Mezei2006, Abele2006}, which have a significantly higher transmission compared to parallel neutron guides and provide a significantly higher neutron flux density at the exit.

For UCNs, similar neutron guides have not yet been designed. Apparently, this is due to the fact that UCN sources based on SF $^4$He have only recently come into use, and only from them can a significant portion of UCNs be extracted through a small hole, since in other sources UCN survive thousands of times less. Some results of calculations of such neutron guides are presented in Section 4, detailed calculations will be presented in a separate publication. 

Let us estimate the density of UCNs in the experimental setup. Let us assume that during the transportation of UCNs along the neutron guide in one pass, half will be lost (calculations show that this is quite possible in the neutron guide described below with a length of $\sim 5~m$, even if it has separating foils and moving devices). Let's also assume that the experimental setup has the same volume as the source and has the same storage time $\tau^0_{stor}$ as it is in the source. In this case $\rho^0_{exp}=\rho_{max}/8$. Indeed, we lost factor 2 in the source, factor 2 in the neutron guide, and factor 2 in the experimental setup. If we exclude $\tau_{He}$ from $\tau^0_{stor}$, since we do not fill the experimental setup with SF $^4$He, then 

\begin{equation}
\rho^0_{exp}=\rho_{max}/6.6=9.2\cdot 10^3~cm^{-3}.
\end{equation}
In this assessment we did not take into account that the UCNs are affected by gravity and that the UCNs must overcome the separating foil before entering the neutron guide. Let us make the corresponding corrections.

First. The exit of the source is located at a height of $\sim$30~cm from the bottom of the source. Not all UCNs produced in the source have sufficient energy to reach it, and the others have reduced their energy by a value corresponding to this height. In fact, this leads to a cutoff of the UCN spectrum on the side of high energies. The calculation for the UCN spectrum (eq. \eqref{eq:spectrum}) in a source with Be walls gives the corresponding correction factor to the UCN output flux: $k_1=0.81$.

Second. With an isotropic distribution of the direction of UCN movement in the source, the flux density of UCNs having energy $E$ at the bottom of the source decreases linearly with increasing height, which leads to a decrease in the exit flux from the source. A calculation gives the corresponding correction factor: $k_2=0.88$.

Third. If a $0.1~mm$ thick separating Al foil is placed behind the diverging mirror cone in the neutron guide, then UCNs will fall on it at an angle close to the normal. In this case, the probability of UCN passing through the foil will be: $k_3=0.74$ for a spectrum (eq. \eqref{eq:spectrum}) in a source with Be walls.

\begin{equation}
    ^{Be}\Phi_{exit}=k_1\cdot k_2\cdot k_3 \cdot ^{Be}\Phi^0_{exit}=0.53\cdot ^{Be}\Phi^0_{exit}.
\end{equation}

The UCN density in the experimental setup will be proportional to the exit flux if the size of the exit hole from the source and the entrance hole into the experimental setup are not changed:

\begin{equation}
    \rho_{exp}=0.53\cdot \rho^0_{exp}=4.9\cdot 10^3~UCN/cm^3.
\end{equation}

When choosing the area of the exit hole from the source, we did not try to optimize it to obtain the maximum value of the flux density, so a higher value can be obtained if needed (compromising the total UCN flux). We also note that the estimated value of the UCN density in the experimental setup would decrease by 25\% if the SF $^4$He temperature in the source is not $\sim$0.9~K, but $\sim$1~K.

Thus, calculations show that if the proposed ideas are feasible, the rate of UCN production can reach $\sim 2.6\cdot 10^7~UCN/s$, the maximum UCN density in the source $\sim 6\cdot 10^4~UCN/cm^3$ and the maximum UCN density in the experimental setup $\sim 5\cdot 10^3~UCN/cm^3$.

\section{Discussion}
\label{sec4}
\subsection{Investigation of the cooling system with a heat conducting wall}
\label{subsec41}

At low temperatures, heat exchange between SF $^4$He and a solid is due to the exchange of energy between phonons at the interface between these media. Due to the large difference in the phonon spectra of the solid and SF $^4$He, this exchange is extremely difficult; phonons of the hotter body are almost completely reflected from the interface. As a result of this effect, a finite temperature difference arises between the solid and SF $^4$He — the Kapitsa temperature jump \cite{Kapitsa1941}, $\Delta T_K$, which is the main obstacle to cooling bodies to ultra-low temperatures. The Kapitsa temperature jump is directly proportional to the heat flux density $Q$ and inversely proportional to $T^3$: $\Delta T_K=R_KQ=AQ/T^3$, where $R_K$ is the Kapitsa resistance. The coefficient $A$ depends on the elasticity of the solid, as well as on the surface roughness and defects of the surface layer, oxides, layers of adsorbed gas, etc.

The temperature difference between SF $^4$He, $\Delta T_{He}$, located on different sides of the heat-conducting wall consists of two Kapitsa temperature jumps at the $^4$He-wall, wall-$^4$He boundaries, and the temperature difference across the wall material. For a given heat flux density through a specific wall, $\Delta T_{He}$ is measured experimentally.

The existence of the Kapitza temperature jump leads to an increase in the SF $^4$He temperature in the UCN source, which contradicts our wish to reduce its temperature. To achieve the required result, we need to increase the efficiency of the $^3$He-$^4$He heat exchanger in order to reduce the temperature of $^4$He in front of the heat-conducting wall of the source. This can be achieved by increasing the effective area of the heat exchanger. In addition, it is necessary to increase the area of the heat-conducting wall in order to reduce the heat flux density through it. We plan to remove heat through the rear wall of the source with a diameter of $\sim$30~cm. With a heat load of $\sim$10~W, the heat flux density through it will be $\sim$140~W/m$^2$. We expect that with this heat flux density $\Delta T_{he}$ will not exceed $\sim$0.2~K at a temperature of $\sim$1~K. If this area is not sufficient, it can be increased several times by also using part of the side wall of the source for heat exchange.

We plan to measure the dependence of $\Delta T_{He}$ on the heat flux density through the heat-conducting wall at a temperature of $\sim$1~K. The measurement is planned to be carried out in a small cryostat with $^3$He vapor pumping, providing a temperature after the heat exchanger of $^4$He$\sim$0.7~K, with a heat inflow of $\sim$0.1~W.

\subsection{Investigation of wall coatings} 
\label{subsec42}

Here, we consider the UCN flux from the source as a function of the critical energy of the source trap walls.

\begin{figure}[t]
    \centering
\includegraphics[width=1.0\columnwidth]{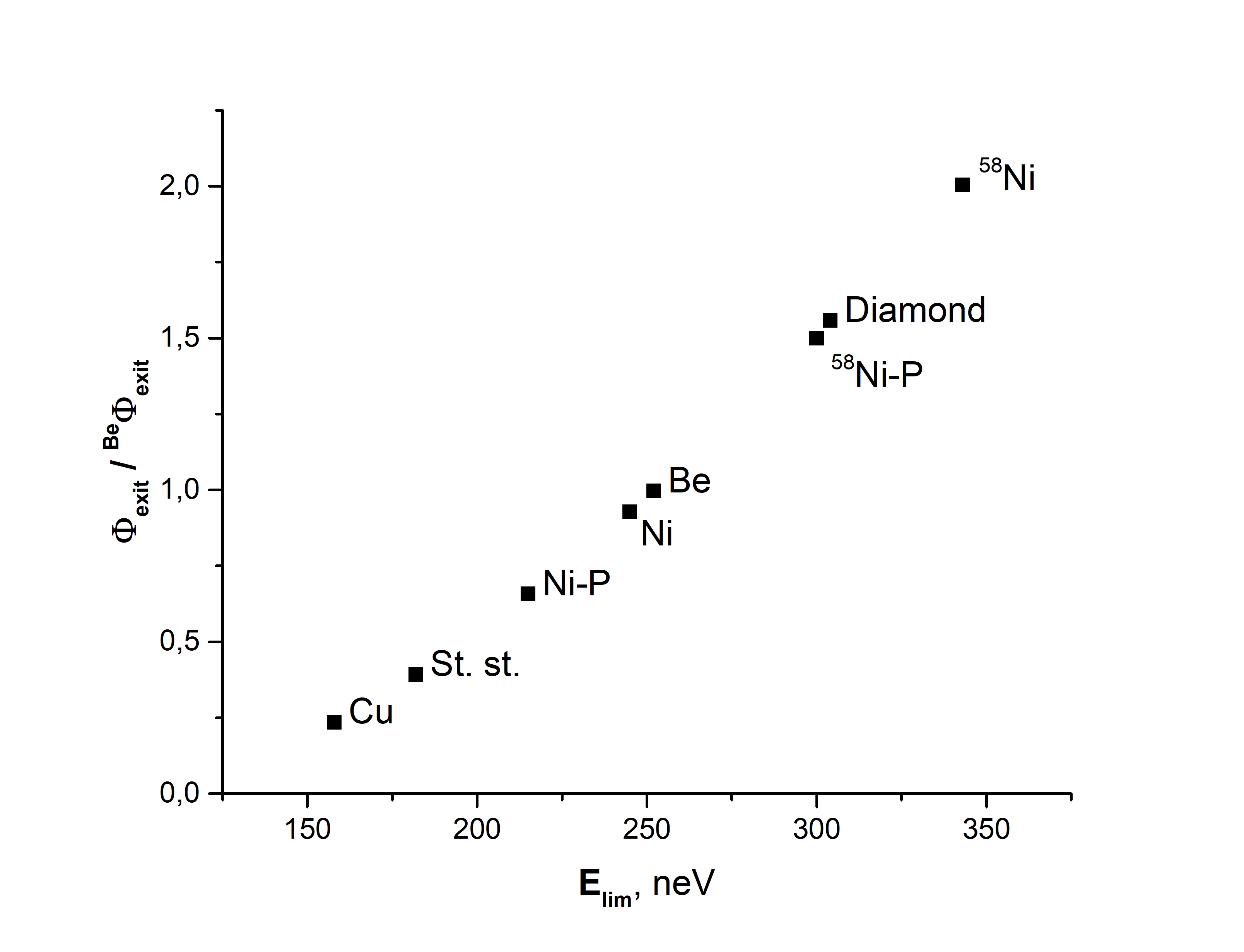}
    \caption{Flux of UCNs from the source for different coatings of its walls, depending on the critical energy of the coating, in relation to the flux from the source with Be walls.}
    \label{fig:DifferentCoatings}
\end{figure}

Fig. \ref{fig:DifferentCoatings} shows the ratio of the UCN flux from the source, $\Phi_{exit}$, for different critical energies of the source walls, to the UCN flux from the source with Be walls, $^{Be}\Phi_{exit}$. This ratio is presented for natural copper (Cu); stainless steel (St. st.); natural nickel-phosphorus coating (Ni-P, phosphorus content 10\%); natural nickel; natural Be; $^{58}$Ni-P coating (phosphorus content 10\%); diamond; and $^{58}$Ni. For all these coatings, the loss factor was assumed to be $\eta =3\cdot 10^{-4}$.

From Fig. \ref{fig:DifferentCoatings} it is clear that the dependence of $\Phi_{exit}$ on the value of the critical energy of the walls is much stronger than $E^{3/2}_{lim}$. If the critical energy increases by 2 times, $\Phi_{exit}$ increases by 7.5 times, and not by 2.8 times. This fact indicates that when manufacturing the source, special attention should be paid to materials with high critical energy.

\begin{figure*}[htbp]
    \centering
\includegraphics[width=0.7\textwidth]{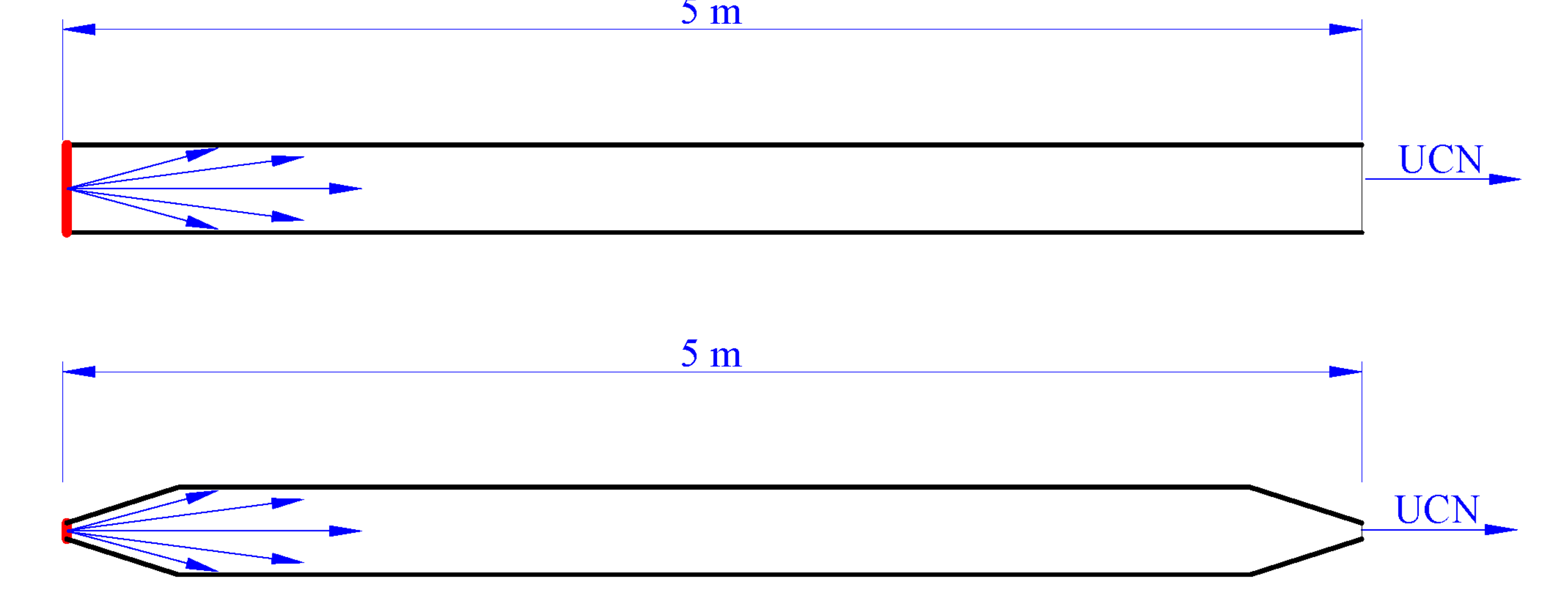}
    \caption{Geometry of the neutron guides for used for calculations. The straight neutron guide is shown at the top, the focusing one at the bottom. UCN sources with Lambert angular distribution are shown in red.}
    \label{fig:NeutronGuides}
\end{figure*}

We are going to carry out a comprehensive study of the coatings with large critical energies, with the purpose of selecting a coating with the maximum critical energy, stable in strong radiation fields at cryogenic temperatures and possessing a minimum UCN loss probability. In particular, we plan to measure the probability of UCN loss, $\mu$, during their interaction with the surfaces of the studied materials, depending on the UCN energy, in the entire energy range from 0 to $E_{lim}$. For this purpose, we plan to irradiate the studied samples with a narrow UCN energy line, with a width of $\Delta E_{UCN}\sim$5–10~neV. Such measurements for high critical energies in the range of $\sim$300~neV have never been performed.

\subsection{Some results of calculations of focusing neutron guides of UCN}
\label{subsec43}

Here we compare the transmission coefficients and flux densities of UCNs at the exit from direct and focusing neutron guides based on numerical simulation using the PENTrack package \cite{Schreyer2017}. 

\begin{figure*}[htbp]
    \centering
    \includegraphics[width=0.95\textwidth]{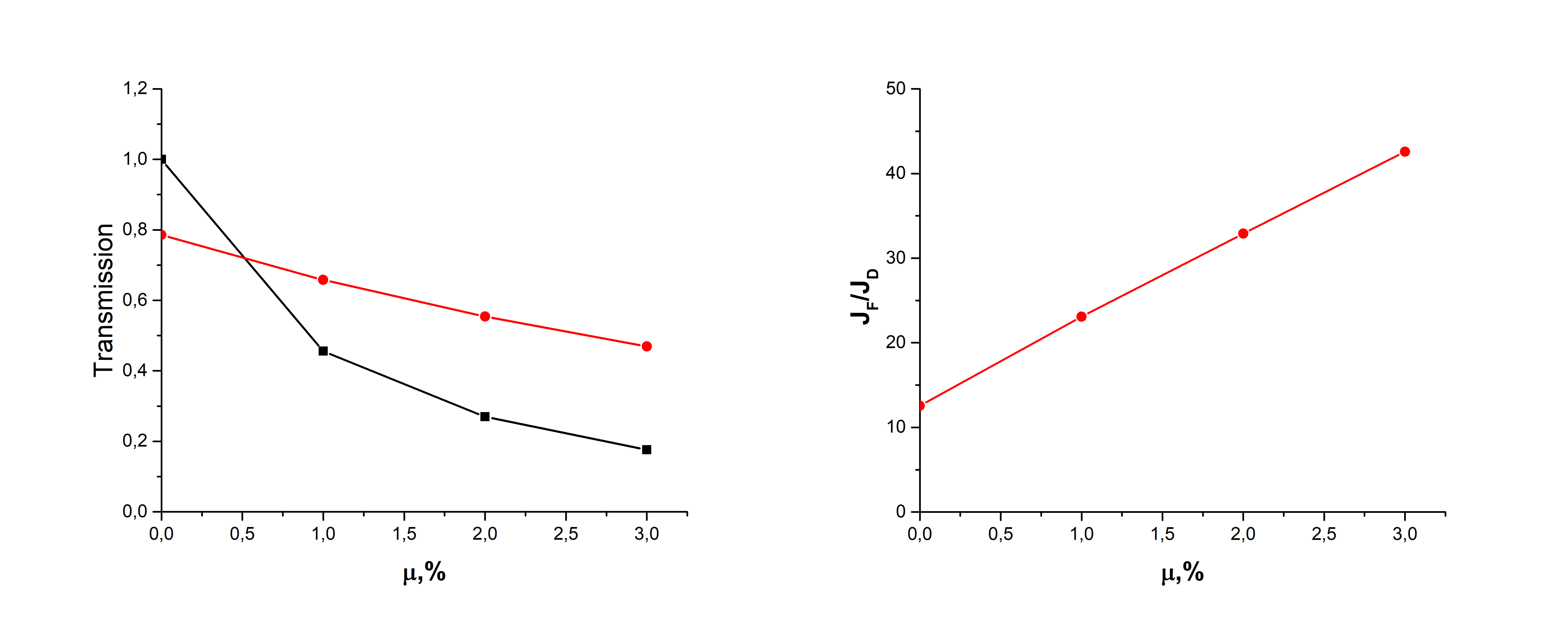}
    
    \caption{On the left are the transmission coefficients of neutron guides depending on the probability of UCN loss, $\mu$, upon impact with the neutron guide wall. The black graph is the transmission of the straight neutron guide; the red graph is the transmission of the focusing neutron guide. On the right is the ratio of the UCN flux density at the exit from the neutron guides. $J_F$ is the flux density at the exit from the focusing neutron guide, $J_D$ is the flux density at the exit from the straight neutron guide.}    \label{fig:TransmissionCoefficients}
\end{figure*}

Fig. \ref{fig:NeutronGuides} shows two geometries of UCN neutron guides. The straight neutron guide has the shape of a cylinder with a diameter of 8~cm and a length of 5~m. The focusing neutron guide consists of an expanding cone, a straight cylindrical neutron guide, and a tapering cone. Each of the two truncated cones has a small diameter of 2~cm, a large diameter of 8~cm, and a length of 30~cm. The diameter of the cylindrical part is 8~cm. The focusing guide total length is 5~m, as in the case of the first geometry. 

Through one end of the neutron guides, marked in red in Fig. \ref{fig:NeutronGuides}, the same UCN flux $F$ enters the neutron guides. The UCNs have the Lambert angular distribution: $d F(\Omega)/d\Omega\sim cos(\theta)$, where $\Omega$ is the solid angle; $\theta$ is the angle between the direction of UCN motion and the normal to the source surface. It corresponds to the isotropic UCN source.

The calculation ignores the UCN losses during their reflection from the guide walls and assumes totally specular reflection. In this case, all UCNs reach the exit of the direct neutron guide. For the focusing neutron guide, 20\% of UCNs are reflected back from the tapering exit part. We assume that all such UCNs are lost. Then the transmission coefficient of the direct neutron guide $T_D=1$, and the focusing guide $T_F=0.8$. 

The mean number of UCN impacts on the walls of the direct neutron guide is  $<ND>=125$, and in the focusing guide it is $<NF>=20$. Fig. \ref{fig:TransmissionCoefficients} (left) shows the transmission coefficients of neutron guides depending on the probability of UCN losses, $\mu$, when hitting the wall and the ratio of UCN flux densities at the exit of the neutron guides (right), $J_F$ is the flux density at the exit from the focusing neutron guide, $J_D$ is the flux density at the exit from the straight neutron guide. As can be seen in this figure, $J_F$ can be ten times greater than $J_D$.

The value of $\mu$ is determined by the probability of off-specular reflection of UCNs from the surfaces of neutron guides. Note that specular reflection of UCNs at any angle of incidence on the surface can be achieved both for well-polished surfaces \cite{Nesvizhevsky2006} and for coatings \cite{Nesvizhevsky2007}. In this case, the value of $\mu$ can be so small that the corresponding additional losses might be only a few percent. Note that the only condition for the partial coefficient of UCN losses due to capture and inelastic scattering is that it is noticeably smaller than the probability of off-specular reflection. This is easily provided even for rectangular neutron guides.

\subsection{Other planned studies}
\label{subsec44}

The idea of designing the UCN source in Almaty, Kazakhstan (AlSUN) was presented and discussed during the dedicated Workshop 'UCN and VCN Source at the Institute of Nuclear Physics', 08-11 April 2024, INP, Almaty, Kazakhstan \cite{INP2024}. This paper is the first description of the main ideas behind this project. The Workshop participants discussed a much wider range of issues related to the concept of the source, the methods for its implementation, and the scientific program using this source. To avoid repetition, we do not reproduce these results here and refer the reader to the materials of this Workshop and the present Special Issue in Materials.

\section{Conclusions}
\label{sect5}

We presented the concept of the UCN source at the WWR-K reactor of the Institute of Nuclear Physics (Almaty, Kazakhstan). The convenient design features of this reactor (the presence of a thermal column, sufficiently high neutron fluxes, and sufficient space to accommodate the source equipment and experimental setups), the existence of well-developed UCN projects of this type (Vancouver and Gatchina), new methods and materials, and a broad international collaboration allow us to expect record parameters of this source. However, some new methods (accumulation and transport of UCNs), as well as the need to cool SF $^4$He to low temperatures under significant thermal load require additional research. This paper can be considered as the first description of the project, showing both its potential and issues requiring additional research.

\section*{Declaration of competing interest}
The authors declare that they have no known competing financial interests or personal relationships that could have appeared to influence the work reported in this paper.

\section*{Funding}

The work was carried out with the financial support of the Science Committee of the Ministry of Science and Higher Education of the Republic of Kazakhstan within the framework of grant funding for young scientists for scientific and scientific-technical projects No. AP19579042.

\section*{CRediT authorship contribution statement}
Conceptualization, \textbf{S.S., K.T., E.K., E.V.L., A.Y.M. and V.V.N.}; methodology, \textbf{K.T., E.K., A.Y.M. and V.V.N.}; software, \textbf{D.S., C.T., P.K.T.}; validation, \textbf{E.K., E.V.L., A.Y.M. and V.V.N.}; 
formal analysis,  \textbf{S.S., K.T., E.K., A.Y.M. and V.V.N.}; investigation, \textbf{S.S., K.T., A.S., D.S., A.S., Zh.K., A.A., O.B., R.K., E.K., E.V.L., A.Y.M., V.V.N., C.T., P.K.T.}; data curation, \textbf{K.T., E.K., E.V.L., A.Y.M. and V.V.N.}; 
writing--original draft preparation, \textbf{K.T., E.K., A.Y.M. and V.V.N.}; writing--review and editing, \textbf{S.S., K.T., A.A., E.K., E.V.L., A.Y.M. and V.V.N.}; visualization, \textbf{K.T., E.K., A.Y.M.}; 
supervision, \textbf{S.S., V.V.N.}; project administration, \textbf{S.S., K.T.}; funding acquisition, \textbf{S.S., K.T.} All authors have read and agreed to the published version of the manuscript.

\section*{Data availability}

The original contributions presented in the study are included in the article, further inquiries can be directed to the corresponding author.

\end{document}